\documentclass[prd,aps,twocolumn,nofootinbib,superscriptaddress,balancelastpage,longbibliography,10pt]{revtex4-2}

\usepackage[T1]{fontenc}
\usepackage[utf8]{inputenc}
\usepackage[english]{babel}
\usepackage{amsmath,amssymb,bm}
\usepackage{graphicx}
\usepackage[dvipsnames]{xcolor}
\usepackage{hyperref}
\hypersetup{colorlinks=true,citecolor=OliveGreen,urlcolor=OliveGreen,linkcolor=OliveGreen}

\newcommand{\Neff}{N_{\rm eff}}
\newcommand{\MZp}{M_{Z'}}
\newcommand{\gprime}{g'}

\newcommand{\dd}{\mathrm{d}}

\newcommand{\optionalfigure}[2]{%
  \IfFileExists{#1}{\includegraphics[width=\linewidth]{#1}}{%
    \fbox{\parbox[c][0.22\textheight][c]{0.92\linewidth}{\centering
    \textit{Figure file \texttt{\detokenize{#1}} not found at compilation time.}\\[3pt]#2}}}}

\newcommand{\optionalfigurewide}[2]{%
  \IfFileExists{#1}{\includegraphics[width=0.85\textwidth]{#1}}{%
    \fbox{\parbox[c][0.25\textheight][c]{0.85\textwidth}{\centering
    \textit{Figure file \texttt{\detokenize{#1}} not found at compilation time.}\\[3pt]#2}}}}

\begin{document}

\title{Constraining $U(1)_{L_\mu-L_\tau}$ Model with Dark Matter Using Relic Density and $N_{\rm eff}$ Bounds}

\author{Shivam Gola}
\email{shivamg.sk@iitb.ac.in}
\affiliation{Department of Physics, Indian Institute of Technology Bombay, Mumbai 400076, India}

\begin{abstract}
We study thermal Dirac dark matter in the gauged $U(1)_{L_\mu-L_\tau}$ model, focusing on resonant annihilation near the $Z'$ pole. We determine the parameter space that satisfies the Planck relic density bound while remaining consistent with collider, astrophysical, and laboratory searches. MeV-scale dark matter and the $Z'$ mediator decouple near the active neutrino decoupling temperature; entropy injection into the neutrino bath modifies $T_\nu/T_\gamma$, leading to a shift in the effective number of relativistic species, $N_{\rm eff}$. Incorporating the coupled temperature evolution alongside the Planck 2018 limit on $N_{\rm eff}$, we find that the viable parameter space is restricted to a narrow mass window of $M_{Z'} \sim 30\text{--}50\text{ MeV}$ with $\gprime \simeq 10^{-7}-10^{-6}$, well within reach of upcoming fixed-target experiments such as SHiP and LDMX.
\end{abstract}

\maketitle

\section{Introduction}
\label{sec:introduction}
Cosmological and astrophysical observations provide compelling evidence for non-baryonic dark matter (DM), while WIMP (Weakly Interacting Massive Particle) produced from thermal freeze-out remains a popular DM candidate~\cite{Profumo:2019ujg,Kolb:1990vq,Bertone:2004pz,Merle:2017jfn,Jungman:1995df,Roszkowski:2017nbc,Schumann_2019}. However, the absence of signals in direct-detection and collider searches places stringent limits on WIMP couplings to quarks~\cite{LUX:2016ggv,XENON:2018voc,Arcadi:2024ukq,CMS:2018mgb}, motivating leptophilic scenarios~\cite{Fox:2008kb,Cohen:2009fz,Bai:2014osa,Chang:2014tea}. A minimal anomaly-free realization is the gauged $U(1)_{L_\mu-L_\tau}$ symmetry~\cite{He:1990pn,He:1991qd,Baek:2001kca}, in which the new gauge mediator $Z'$ couples at tree level to second- and third-generation leptons, while couplings to electrons and quarks arise only through loop-induced kinetic mixing~\cite{Holdom:1985ag,Lindner:2016bgg,Araki:2017wyg,Escudero:2019gzq}. We also introduce a Dirac fermion DM candidate $\chi$ with unit charge under the new gauge group; as a result, it evades tree-level hadronic direct-detection bounds~\cite{Baek:2008nz,Patra:2016shz,Foldenauer:2018zrz,Holst:2021lzm,Hapitas:2021ilr,Figueroa:2024tmn}.

The light $Z'$ mediator is constrained by laboratory and astrophysical searches, including NA64$\mu$~\cite{NA64:2024klw}, BABAR~\cite{BaBar:2016sci}, and white-dwarf cooling~\cite{Foldenauer:2024cdp}, etc. The $U(1)_{L_\mu-L_\tau}$ extension has also been extensively studied in relation to the muon anomalous magnetic moment $(g-2)_\mu$; however, previous studies show that favorable couplings lie around $\gprime \sim 4\times 10^{-4}\text{--}10^{-3}$ for sub-GeV mediators~\cite{Baek:2001kca,Lindner:2016bgg,Drees:2021rsg,Baek:2022ozm}. These constraints motivate resonant annihilation as a viable alternative~\cite{Griest:1990kh,Ibe:2008ye,Guo:2009aj,Foldenauer:2018zrz}. We define a positive tuning parameter $\delta$ in Eq.~\ref{eq:tuning} such that $\delta \to 0$ corresponds to approaching the $Z'$ pole. The decay channel $Z'\to\chi\bar\chi$ is open for $\delta < 1$. In this regime, resonant enhancement can reproduce the observed relic abundance for much smaller $\gprime$ than would be required away from the pole. In particular, for a benchmark tuning of $\delta = 10^{-2}$, the thermally averaged annihilation cross section is enhanced sufficiently that the observed relic abundance is obtained for $\gprime \sim 10^{-7}\text{--}10^{-6}$, several orders of magnitude below the $(g-2)_\mu$ band and current laboratory exclusions.

For $\gprime \gtrsim 10^{-8}$, rapid $Z' \leftrightarrow \nu_{\mu,\tau}\bar\nu_{\mu,\tau}$ and $Z' \leftrightarrow \chi\bar\chi$ reactions thermalize the dark sector with the Standard Model (SM) neutrino bath in the early Universe~\cite{Escudero:2019gzq}, allowing the resonant annihilation rate and equilibrium density to be evaluated at the neutrino temperature $T_\nu(T_\gamma)$. Additionally, in the sub-GeV regime, thermal $\chi$ and $Z'$ populations can exchange entropy with leptons around neutrino decoupling $T\sim 2\,\mathrm{MeV}$, modifying $T_\gamma, ~T_\nu$, and therefore the effective number of relativistic species, $\Neff$~\cite{Escudero:2019gzq,Sabti:2019mhn,Hambye:2021moy}. Since the resulting shift depends on how entropy is shared between neutrinos and charged leptons~\cite{Escudero:2018mvt,EscuderoAbenza:2020cmq}, $N_{\rm eff}$ provides a complementary probe. In this work, we study the resonant production of DM and its cosmological impact using coupled temperature evolution equations for the electromagnetic and neutrino baths. We compare the resulting cosmological constraints with laboratory, collider, and astrophysical bounds in the $(\MZp,\gprime)$ plane.

The paper is organized as follows. Section~\ref{sec:model} presents the model, and Section~\ref{sec:relic} discusses thermal freeze-out. Section~\ref{sec:neff} develops the $\Neff$ methodology. Section~\ref{sec:results} summarizes our main results and the constraints, and Section~\ref{sec:conclusion} concludes.

\section{The Model}
\label{sec:model}

We consider a gauged $U(1)_{L_\mu-L_\tau} $ model and a SM gauge singlet Dirac fermion $\chi$ with unit charge under the new gauge group. A $Z_2$ symmetry is imposed under which $\chi$ is odd, while all SM fields and $Z'$ are even; this forbids mixing with $
\nu_\mu$ and stabilizes $\chi$. We treat the DM and $Z'$ mass $m_\chi$ and $\MZp$ respectively as independent low-energy parameters and do not specify the ultraviolet origin of these masses. The interactions relevant for annihilation and neutrino-decoupling thermodynamics are \cite{Escudero:2019gzq,Foldenauer:2018zrz,Wang:2025lmtDM}:
\begin{align}
\mathcal{L}_{\rm int}\supset \gprime Z'_\alpha\big(&\bar\mu\gamma^\alpha\mu-\bar\tau\gamma^\alpha\tau
+\bar\nu_\mu\gamma^\alpha P_L\nu_\mu-\bar\nu_\tau\gamma^\alpha P_L\nu_\tau \nonumber\\
&+\bar\chi\gamma^\alpha\chi\big).
\label{eq:lagrangian}
\end{align}
Here, $P_{L,R}=(1\pm\gamma_5)/2$ denotes the left(right)-handed projection operator, and $\gprime$ is the new gauge coupling. We set the tree-level kinetic mixing to zero; however the loop diagrams induce the photon--$Z'$ mixing~\cite{Holdom:1985ag,Kamada:2015era,Araki:2017wyg},
\begin{align}
\mathcal{L}_{\rm mix}&=-\frac{\varepsilon}{2}F_{\alpha\beta}F'^{\alpha\beta}.
\label{eq:mixing}
\end{align}
For low momentum transfer \cite{Escudero:2019gzq,Hapitas:2021ilr,Wang:2025lmtDM}, the loop-induced mixing $\varepsilon\simeq-\frac{\gprime}{70}$, which induces the electron coupling $g_e=\varepsilon e$. For a kinematically accessible fermion pair $f\bar f$ in our model, the partial width \cite{Foldenauer:2018zrz} is
\begin{align}
\Gamma_{Z'\to f\bar f}=
\frac{\eta_f g_f^2\MZp}{12\pi}
\left(1+\frac{2m_f^2}{\MZp^2}\right)
\sqrt{1-\frac{4m_f^2}{\MZp^2}}.
\label{eq:widthfermion}
\end{align}
where $g_\chi = g_f = g'$ for all fermions except electrons, $g_f = \varepsilon e$; $\eta_f = 1/2~(1)$ for neutrino flavor (for a charged fermion). The total width $\Gamma_{Z'}$ is the sum over all kinematically allowed channels.

\section{Relic density}
\label{sec:relic}

The number density of thermal dark matter, $n_\chi$, evolves according to the Boltzmann equation \cite{Kolb:1990vq,Gondolo:1990dk}:
\begin{align}
\frac{\dd n_\chi}{\dd t} + 3Hn_\chi = -\langle\sigma v\rangle
\left(n_\chi^2 - (n^{\text{eq}}_{\chi})^2\right),
\label{eq:boltzmann}
\end{align}
where $H \equiv H(T) = \sqrt{\frac{\pi^2 g_*(T)}{90}} \frac{T^2}{M_{\text{Pl}}}$ is the Hubble parameter, $g_{*}(T)$ is the effective number of degrees of freedom in the thermal bath~\cite{Kolb:1990vq}, $M_{\text{Pl}} \simeq 2.4 \times 10^{18} \, \text{GeV}$ and $n^{\text{eq}}_{\chi}$ denotes the equilibrium number density of $\chi$.

Annihilation proceeds via an $s$-channel mediator into kinematically accessible SM fermions, including the loop-induced $e^+e^-$ channel. The cross section for the annihilation channel $\chi \bar\chi \rightarrow f\bar f$ \cite{Foldenauer:2018zrz} is given by
\begin{align}
\sigma_f(s) = {}&\frac{\eta_f g_\chi^2 g_f^2}{12\pi s}
\sqrt{\frac{s - 4m_f^2}{s - 4m_\chi^2}}
\frac{(s + 2m_\chi^2)(s + 2m_f^2)}
{(s - M_{Z'}^2)^2 + M_{Z'}^2\Gamma_{Z'}^2},
\label{eq:sigmafull}
\end{align}
Using Maxwell--Boltzmann (MB) statistics, the thermally averaged cross section at the dark matter temperature $T_\chi$ is expressed as \cite{Gondolo:1990dk}
\begin{equation}
\begin{split}
\langle\sigma v\rangle(T_\chi) = {}&\frac{1}{8m_\chi^4 T_\chi K_2^2(m_\chi/T_\chi)}\\
&\times\int_{4m_\chi^2}^{\infty}\!\dd s\,\sigma(s)(s - 4m_\chi^2)\sqrt{s}\,
K_1\!\left(\frac{\sqrt{s}}{T_\chi}\right),
\end{split}
\label{eq:thermalaverage}
\end{equation}
with $\sigma(s) = \sum_f \sigma_f(s)$. Since DM interacts more strongly with neutrinos than with electrons in our model, it follows the neutrino temperature, $T_\chi = T_\nu(T_\gamma)$~\cite{Escudero:2019gzq}, which we determine in Sec.~\ref{sec:neff}. 

To investigate resonant annihilation, which occurs when the dark matter mass lies near the $Z'$ pole, we parameterize this mass tuning using a dimensionless parameter $\delta$ by
\begin{align}
m_\chi = \frac{M_{Z'}}{2}\sqrt{1 - \delta} \ , \quad \ 0 < \delta < 1. 
\label{eq:tuning}
\end{align}
We solve Eq.~\eqref{eq:boltzmann} numerically in terms of the comoving yield $y_\chi \equiv n_\chi / s_{\rm SM}(T_\gamma)$, where, $s_{\rm SM}(T_\gamma) = \frac{2\pi^2}{45} g_{*s}(T_\gamma) T_\gamma^3$ represents the SM entropy density, $g_{*s}$ is the effective degree of freedom~\cite{Kolb:1990vq}. Using the dimensionless variable $x \equiv m_\chi / T_\chi$, the present-day relic abundance is subsequently extracted from the asymptotic yield $y_\chi(x_\infty)$ via $\Omega_\chi h^2 = m_\chi y_\chi(x_\infty) s_0 h^2 / \rho_c$, where $s_0 \simeq 2891.2 \, \text{cm}^{-3}$ is the present-day entropy density and $\rho_c \simeq 1.05 \times 10^{-5} h^2 \, \text{GeV} \, \text{cm}^{-3}$ is the critical density of the Universe. Finally, we identify the parameter space consistent with the Planck $3\sigma$ relic-density band, $\Omega_\chi h^2 = 0.1200 \pm 0.001$ \cite{Planck:2018vyg}.

\section{Neutrino decoupling}
\label{sec:neff}

At a few MeV temperature, active neutrinos decouple from the electromagnetic plasma and evolve into three free-streaming species, implying a finite neutrino energy density which is parameterized by $N_{\text{eff}}$:
\begin{align}
N_{\text{eff}} = 3\left(\frac{11}{4}\right)^{4/3}\left(\frac{T_\nu}{T_\gamma}\right)^4.
\label{eq:neff}
\end{align}
In SM, precision calculations yield $N_{\rm eff}^{\rm SM} = 3.044$ \cite{Akita:2020szl,Froustey:2020mcq,Bennett:2020zkv}. However, the Planck 2018 collaboration reports $N_{\text{eff}} = 2.99 \pm 0.17$ from their data fit, corresponding to an approximate $1\sigma$ upper limit, $N_{\text{eff}} \leq 3.16$ \cite{Planck:2018vyg}. To estimate $N_{\text{eff}}$ in our model, we follow the prescriptions of Refs.~\cite{Escudero:2018mvt,Escudero:2019gzq,EscuderoAbenza:2020cmq,Sabti:2019mhn}, assuming that neutrino oscillations maintain a common neutrino temperature, $T_{\nu_i} = T_\nu$, where $i=e,\mu,\tau$ \cite{Hannestad:2001iy}. For $g' \gtrsim 10^{-8}$, rapid $Z' \leftrightarrow \nu_{\mu,\tau}\bar\nu_{\mu,\tau}$ and $Z' \leftrightarrow \chi\bar\chi$ processes ensure that both the $Z'$ mediator and the dark matter particle $\chi$ remain fully thermalized with the neutrino bath ($T_{Z'} = T_\chi = T_\nu$) in the early Universe~\cite{Escudero:2019gzq}. Though DM shares the neutrino temperature, a non-negligible DM-electron energy transfer rate can modify the photon temperature, which we quantify by $\dfrac{\delta\rho_{\chi-e}}{\delta t}$~\cite{Escudero:2019gzq}. To determine the evolution of $T_\nu$ and $T_\gamma$, we write down the following coupled Boltzmann equations, using the method described in \cite{Escudero:2018mvt,Escudero:2019gzq,EscuderoAbenza:2020cmq}:
\begin{widetext}
\begin{align}
\frac{dT_\nu}{dt} &= -
\frac{4H\rho_\nu + 3H(\rho_{Z'} + P_{Z'}) + 3H(\rho_\chi + P_\chi)
-\dfrac{\delta\rho_{\nu-e}}{\delta t}-\dfrac{\delta\rho_{Z'-e}}{\delta t}-\dfrac{\delta\rho_{\chi-e}}{\delta t}}
{\dfrac{\partial\rho_\nu}{\partial T_\nu}+\dfrac{\partial\rho_{Z'}}{\partial T_\nu}+\dfrac{\partial\rho_\chi}{\partial T_\nu}},
\label{eq:nux}\\
\frac{dT_\gamma}{dt} &= -
\frac{4H\rho_\gamma + 3H(\rho_e + P_e)
+\dfrac{\delta\rho_{\nu-e}}{\delta t}+\dfrac{\delta\rho_{Z'-e}}{\delta t}+\dfrac{\delta\rho_{\chi-e}}{\delta t} }
{\dfrac{\partial\rho_\gamma}{\partial T_\gamma}+\dfrac{\partial\rho_e}{\partial T_\gamma}}.
\label{eq:CBE}
\end{align}
\end{widetext}
Here, $\rho_i$ and $P_i$ denote the energy density and pressure of species $i$, evaluated using Bose--Einstein or Fermi--Dirac distribution functions according to particle statistics. The transfer rate $\delta\rho_{a-b}/\delta t$ describes the energy exchange between particles $a$ and $b$~\cite{Escudero:2018mvt,EscuderoAbenza:2020cmq}; we list these rates for all participating species in the neutrino decoupling process in Appendix~\ref{app:rates}. We initialize the system at $T_\nu(t_{\rm eq}) = T_\gamma(t_{\rm eq}) \approx \text{max}\{10~\text{MeV}, ~M_{Z'}\} $ with $t_{\rm eq} = 1/2H(T_{\max})$. Finally, Eq.~\eqref{eq:nux}--\eqref{eq:CBE} are integrated to a sufficiently large final time $t_f$ by which $e^+ e^-$ annihilation is complete.

\section{Results}
\label{sec:results}

\begin{figure}[htbp]
    \centering
    \includegraphics[width=0.49\textwidth]{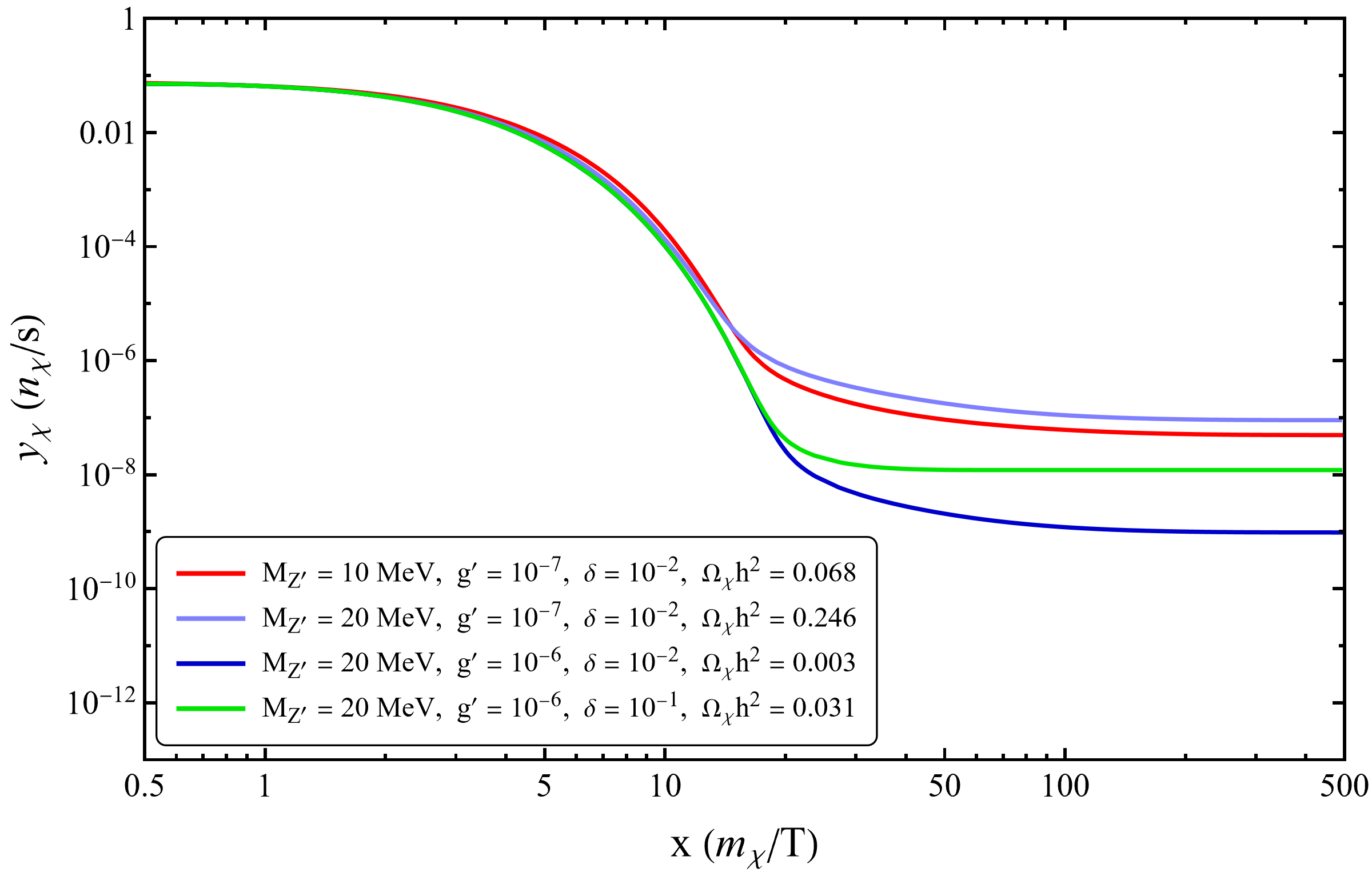}
    \caption{Evolution of the comoving dark-matter yield $y_\chi$ as a function of $x$ for four benchmarks.}
    \label{fig:relic}
\end{figure}

In this section, we present our analysis and results. The yield curves in Fig.~\ref{fig:relic} illustrate the evolution of the comoving DM abundance for the four benchmark points. Near the resonance, annihilations remain efficient well beyond $x \sim 20$, and the yield approaches its asymptotic plateau only by $x \sim 100$ for $\delta = 10^{-2}$. The red and light-blue curves are at fixed $\gprime = 10^{-7}$ and $\delta = 10^{-2}$, but $\MZp$ is doubled. Since the annihilation cross section scales as the inverse square of $\MZp$, freeze-out occurs earlier for $\MZp=20$ MeV and therefore results in larger relic abundances. Increasing the gauge coupling enhances the thermally averaged annihilation cross section, which results in smaller DM relic abundances; this is reflected in the benchmark shown by the light-blue and dark-blue curves. Lastly, increasing the tuning parameter $\delta = 10^{-1}$ while other parameters are kept the same as the dark-blue curve implies a decrease in the thermal cross-section; therefore, freeze-out occurs earlier and results in a higher relic abundance, as shown by comparing the dark-blue and green curves.

\begin{figure}[htbp]
    \centering
    \includegraphics[width=0.49\textwidth]{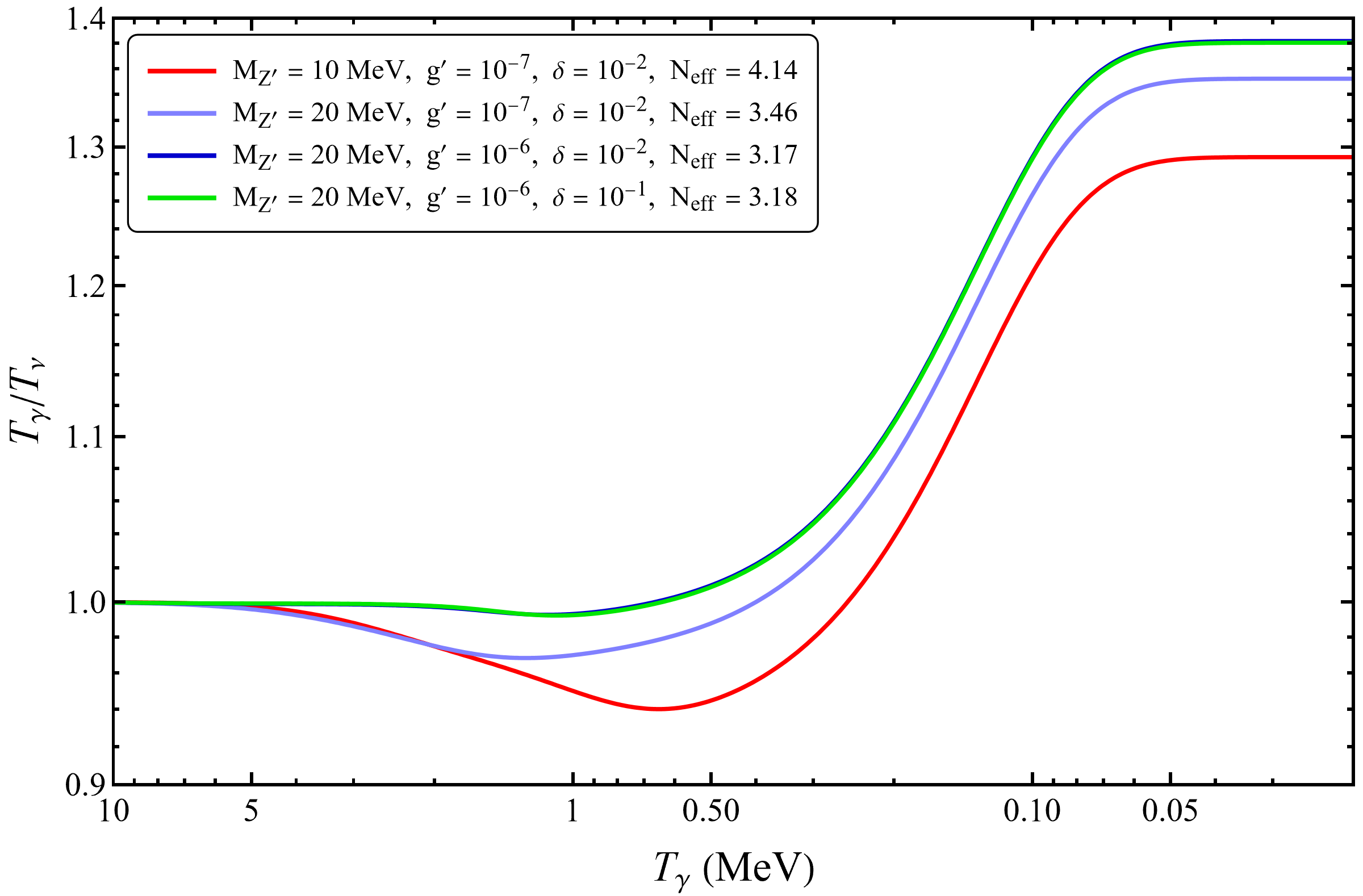}
    \caption{Evolution of $T_\gamma/T_\nu$ versus photon temperature $T_\gamma$ for the same benchmarks.}
    \label{fig:neff}
\end{figure}

\begin{figure*}[htbp]
\centering
\includegraphics[width=\textwidth]{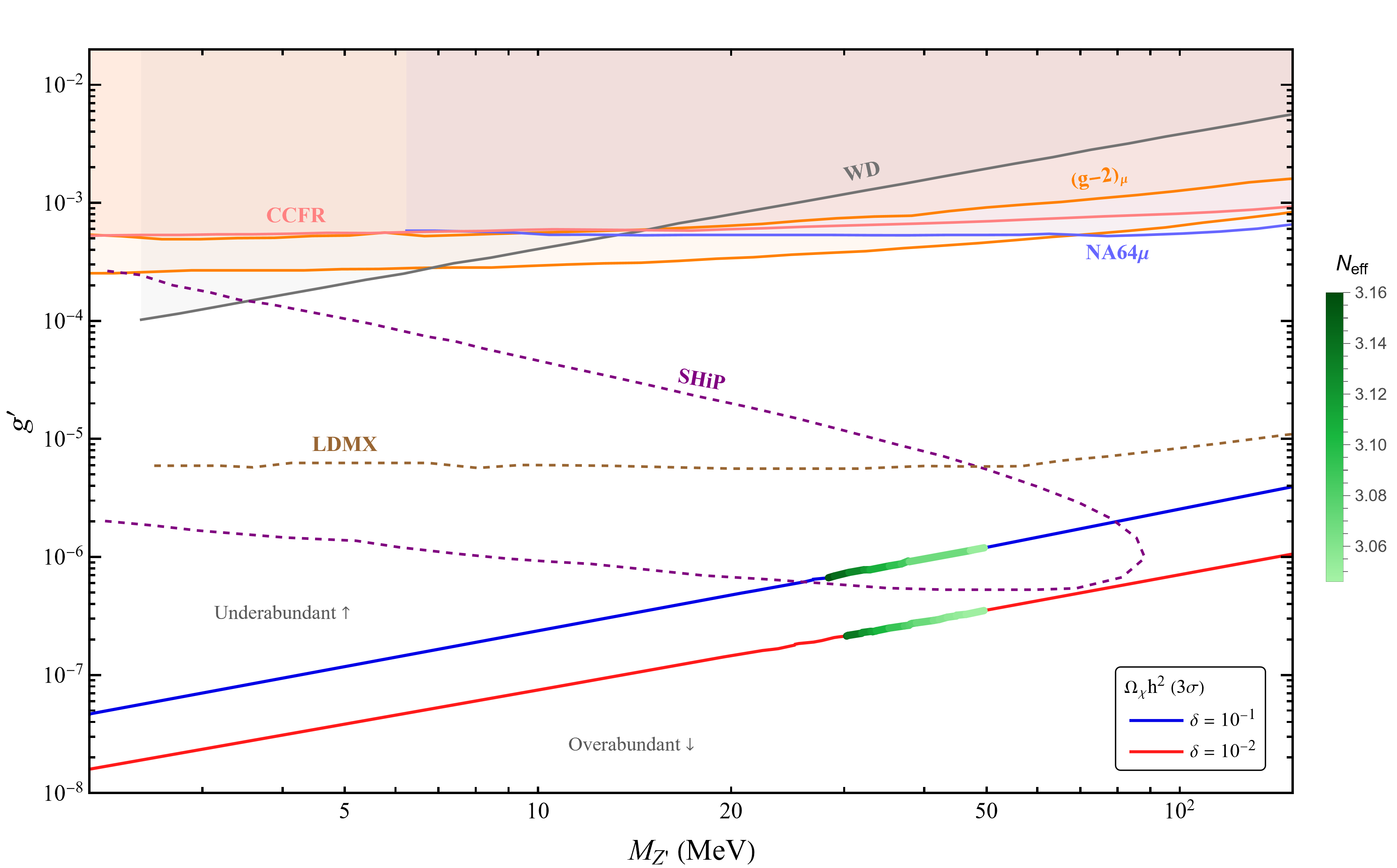}
\caption{Constraints and the allowed parameter space are shown in the $(M_{Z'},g')$ plane . The contours from our model are displayed in blue and red, representing the $3\sigma$ Planck relic-density bound, while the green shade overlay satisfies the $1\sigma$ Planck upper bound on $\Neff \leq 3.16$ for $\delta=10^{-1,-2}$, respectively. The experimental constraints include CCFR~\cite{CCFR:1991trident,Altmannshofer:2014pba}, NA64$\mu$~\cite{NA64:2024klw}, white-dwarf (WD) cooling~\cite{Foldenauer:2024cdp}, $(g-2)_\mu$~\cite{Muong-2:2021ojo,Muong-2:2023cdq}, together with the projected sensitivities of LDMX~\cite{Kahn:2018cqs,Berlin:2018bsc}, and, SHiP~\cite{Alekhin:2015byh}.}
\label{fig:constraints}
\end{figure*}

In Fig.~\ref{fig:neff}, we show the evolution of $T_\gamma/T_\nu$ with photon temperature $T_\gamma$ for the same benchmark points as in Fig.~\ref{fig:relic}. At high temperature $T \approx \MZp$, all species share a common temperature, $T_\gamma=T_\nu$. Below $T\sim2$--$3\,\mathrm{MeV}$, weak interactions freeze out and $T_\gamma \neq T_\nu$ anymore, however, $Z'$ and $\chi$ remain coupled to neutrinos through rapid $Z'\leftrightarrow\nu\bar\nu$ and $Z' \leftrightarrow  \chi\bar\chi$ processes. As $T_\nu$ falls below $\MZp$ and $m_\chi$, these species become non-relativistic and transfer their entropy to neutrinos, so $T_\gamma/T_\nu$ drops around $T_\gamma\sim0.5$. Afterward, $e^+e^-$ annihilation heats the photons and $T_\gamma/T_\nu$ approaches its asymptotic value; this modifies the $\Neff$. The effect is driven mainly by the dark-sector masses: lighter dark-sector particles~(DM and $\MZp$) release more entropy after neutrino decoupling and give a sharper dip effect and larger $\Neff$ as displayed by the benchmark red and light-blue curves. However, increasing the gauge coupling makes $Z'\leftrightarrow e^+e^-$ efficient below the neutrino decoupling temperature, returning some entropy to the photon bath, which produces a shallower dip and a smaller $\Neff$ as shown by the light- and dark-blue color benchmark curves. The nearly identical dark-blue and green curves correspond to $\delta = 10^{-2}$ and $10^{-1}$, respectively, while all other benchmark parameters are kept the same. The curves show that $N_{\rm eff}$ is largely insensitive to the tuning parameter $\delta$, since $m_\chi$ does not change significantly for the given $\delta$ values in the benchmarks. In addition, the transfer rate for the $\chi$--$e$ process is suppressed by $(\varepsilon e)^2$.

In Figure~\ref{fig:constraints}, we determine the viable parameter space in the $(\MZp, \gprime)$ plane by imposing the DM relic density bound, $\Neff$, and constraints from laboratory, collider, and astrophysical searches. We include exclusion limits from missing-energy searches from NA64$\mu$~\cite{NA64:2024klw}, white-dwarf (WD) cooling~\cite{Foldenauer:2024cdp}, $(g-2)_\mu$ measurement bounds~\cite{Muong-2:2021ojo,Muong-2:2023cdq}, and neutrino trident at CCFR~\cite{CCFR:1991trident}, additionally, we display projected sensitivities from the LDMX fixed-target experiment~\cite{Kahn:2018cqs,Berlin:2018bsc} and the SHiP beam-dump facility~\cite{Alekhin:2015byh}. We solve the coupled Boltzmann Eqs.~\eqref{eq:boltzmann} and \eqref{eq:nux}--\eqref{eq:CBE} numerically and evaluate the DM yield and neutrino decoupling across the parameter space. The solid blue and red contours correspond to the Planck $3\sigma$ dark matter relic abundance for $\delta = 10^{-1}$ and $\delta = 10^{-2}$, respectively. The smaller $\delta$ accommodates lower values of the gauge coupling $\gprime$, as expected from resonant enhancement. Parameter regions above the contours yield a dark matter underabundance, whereas those below yield an overabundance. The green shaded color overlay on the $\delta = 10^{-1,-2}$ contours indicates the shift in $\Neff$, with the Planck $1\sigma$ upper bound $\Neff \leq 3.16$. This identifies an allowed parameter window around $\MZp \simeq 30\text{--}50\,\mathrm{MeV}$ with $\gprime \simeq (2.1\text{--}3.5)\times 10^{-7}$ for $\delta = 10^{-2}$, which evades all existing laboratory and astrophysical constraints, while future experiments at SHiP and LDMX could probe this region. However, the green shaded overlay for $\delta = 10^{-1}$ shows that $\gprime \simeq (6.6\text{--}12)\times 10^{-7}$ with $\MZp \simeq 30\text{--}50\,\mathrm{MeV}$ lies within the current sensitivity of SHiP. Furthermore, direct-detection constraints remain strongly suppressed in our scenario. The $Z'$ couples to second- and third-generation leptons at tree level; however, couplings to electrons and quarks arise exclusively through loop-induced kinetic mixing. For benchmark values $\gprime \sim 10^{-7}$, $\MZp \sim 30\,\mathrm{MeV}$, and $m_\chi \sim \MZp/2$, the resulting spin-independent DM--nucleon cross section for Xe nuclei is estimated as $\sigma_{\chi N} \sim \mathcal{O}(10^{-60})\,\mathrm{cm}^2$~\cite{Wang:2025lmtDM,Foldenauer:2018zrz}, which is orders of magnitude below current direct-detection thresholds~\cite{abdelhameed2019first}.

\section{Conclusion}
\label{sec:conclusion}

We investigated resonant thermal Dirac dark matter in the gauged $U(1)_{L_\mu-L_\tau}$ framework and assessed its effect on neutrino decoupling. The DM relic density is set by an enhanced resonant annihilation cross-section near the $Z'$ pole, while the temperature evolution follows the entropy deposited into the neutrino bath relative to the electromagnetic plasma. We compared the resulting cosmological constraints with existing laboratory, collider, and astrophysical bounds, along with projected sensitivities. Our benchmark evolutions in Figs.~\ref{fig:relic} and \ref{fig:neff} show that the mediator mass $\MZp$, gauge coupling $\gprime$, and mass-tuning parameter $\delta$ jointly determine the thermal yield and the neutrino temperature after decoupling. We present our main result in Fig.~\ref{fig:constraints} in the $(\MZp, \gprime)$ plane as the contours satisfying the Planck $3\sigma$ relic-density bound for $\delta = 10^{-1,-2}$. Furthermore, the Planck $1\sigma$ upper bound $\Neff \leq 3.16$, isolates a small viable parameter window around $\MZp \simeq 30\text{--}50\,\mathrm{MeV}$ ($m_\chi \simeq 15\text{--}25\,\mathrm{MeV}$) with $\gprime \simeq (2.1\text{--}3.5)\times 10^{-7}$ along the $\delta = 10^{-2}$ contour. This allowed region evades current laboratory exclusions, collider limits, and white-dwarf cooling constraints, while the allowed region for $\delta = 10^{-1}$ lies within the reach of current SHiP sensitivities; the allowed region for $\delta = 10^{-2}$ remains fully accessible to prospective searches at LDMX~\cite{Kahn:2018cqs,Berlin:2018bsc} and SHiP~\cite{Alekhin:2015byh}. A full phase-space treatment of the resonant cross-section at smaller $\delta$, where early kinetic decoupling effects become significant~\cite{Duch:2017nbe,Binder:2017rgn, Binder:2021bmg}, remains a promising direction for future work.

\section*{Acknowledgments}
We thank Debasish Borah, Manibrata Sen, and Rahul Srivastava for helpful discussions and insightful comments.

\appendix

\section{Thermodynamic Formulae}
\label{app:thermo}

For a particle of mass $m$, internal degrees of freedom $g$, and temperature $T$, the equilibrium number density, energy density, and pressure are~\cite{Kolb:1990vq}
\begin{align}
n_\text{eq}&=\frac{g}{2\pi^2}\int_m^\infty\dd E\,
\frac{E\sqrt{E^2-m^2}}{\exp(E/T)\pm1}, \\
\rho_\text{eq}&=\frac{g}{2\pi^2}\int_m^\infty\dd E\,
\frac{E^2\sqrt{E^2-m^2}}{\exp(E/T)\pm1}, \\
P_\text{eq}&=\frac{g}{6\pi^2}\int_m^\infty\dd E\,
\frac{(E^2-m^2)^{3/2}}{\exp(E/T)\pm1},
\label{eq:thermo}
\end{align}
where $+1$ applies to fermions and $-1$ to bosons. In the Maxwell--Boltzmann approximation,
\begin{align}
n_{\rm eq}&=g\frac{m^2T}{2\pi^2}K_2(m/T), \\
\rho_{\rm eq}&=g\frac{m^2T}{2\pi^2}\left[mK_1(m/T)+3TK_2(m/T)\right], \\
P_{\rm eq}&=g\frac{m^2T^2}{2\pi^2}K_2(m/T),
\label{eq:thermoMB}
\end{align}
where $K_{1,2}$ are the modified Bessel functions.

\section{The Boltzmann equation}
\label{app:BE}

To track the temperature evolution of a species $X$, we integrate the Liouville equation over momentum \cite{Kolb:1990vq}:
\begin{align}
\frac{\partial f_X(p,t)}{\partial t} - H p \frac{\partial f_X(p,t)}{\partial p} = \mathcal{C}[f_X] \,,
\label{eq:BE1}
\end{align}
where $f_X(p,t)$ is the distribution function and $\mathcal{C}[f_X]$ is the collision term. For a $2\to2$ process $X+i \to j+k$, assuming CP invariance, the collision term is
\begin{align}
\mathcal{C}[f_X] \equiv{}& - \frac{1}{2 E_X} \int \dd\Pi_{i}\, \dd\Pi_{j}\, \dd\Pi_{k}\, (2\pi)^4 \delta^{(4)}(p_X + p_i - p_j - p_k) \nonumber \\
& \times |\mathcal{M}|^{2}_{X+i \to j+k} \bigl[ f_X f_{i} (1 \pm f_j)(1 \pm f_k) \nonumber \\
& - f_{j} f_k (1 \pm f_X)(1 \pm f_i) \bigr] \,,
\label{eq:cf}
\end{align}
where $|\mathcal{M}|^2$ is the spin-averaged squared matrix element, $\dd\Pi_a=\dd^3p_a/[(2\pi)^3 2E_a]$ is the invariant phase-space measure, and the upper (lower) sign applies to bosons (fermions). Multiplying Eq.~\eqref{eq:BE1} by $g_X E_X \dd^3 p_X / (2\pi)^3$ and integrating gives the energy-density evolution equation:
\begin{align}
\frac{\dd\rho_X}{\dd t} + 3 H (\rho_X + P_X) = \frac{\delta \rho_{X}}{\delta t} \,,
\label{eq:Etrans}
\end{align}
where $\frac{\delta \rho_X}{\delta t}$ is the net energy transfer rate. Rewriting in terms of temperature~\cite{Escudero:2018mvt,EscuderoAbenza:2020cmq}, this becomes
\begin{align}
\frac{\dd T_{X}}{\dd t} = \left( - 3 H (\rho_{X}+P_X) + \frac{\delta \rho_{X}}{\delta t} \right) \left( \frac{\partial\rho_{X}}{\partial T_{X}} \right)^{-1} \,.
\label{eq:dTXdt}
\end{align}
Eq.~\eqref{eq:dTXdt} is utilized to write the coupled temperature evolution equations for neutrino, electron, photon, and the dark-sector species.

\begin{table*}[htbp]
\begin{ruledtabular}
\begin{tabular}{ccc} 
Process & $2^{-5} G_F^2 \overline{|\mathcal{M}|^{2}}$ & $\pi^5 G_F^{-2} \frac{\delta \rho}{\delta t}$ \\ \hline
$\nu_i + \bar{\nu}_i \leftrightarrow e^+ + e^-$ & $4[g_{L_i}^2(p_1\!\cdot\! p_4)(p_2\!\cdot\! p_3)+g_{R_i}^2(p_1\!\cdot\! p_3)(p_2\!\cdot\! p_4)]$ & $64 (g_{L_i}^2+g_{R_i}^2) \bigl[ T_{\gamma}^9 - T_{\nu_i}^9 \bigr]$ \\[4pt] 
$\nu_i + e^{\pm} \leftrightarrow \nu_i + e^{\pm}$ & $4[g_{L_i}^2(p_1\!\cdot\! p_2)(p_3\!\cdot\! p_4)+g_{R_i}^2(p_1\!\cdot\! p_4)(p_2\!\cdot\! p_3)]$ & $112(g_{L_i}^2+g_{R_i}^2)\, T_{\gamma}^4 T_{\nu_i}^4 \bigl[ T_{\gamma} - T_{\nu_i} \bigr]$ \\[4pt]
$\nu_i + \nu_j \leftrightarrow \nu_i + \nu_j$ & $(p_1\!\cdot\! p_2)(p_3\!\cdot\! p_4)$ & $48\, T_{\nu_j}^4 T_{\nu_i}^4 \bigl[ T_{\nu_j} - T_{\nu_i} \bigr]$ \\[4pt]
$\nu_i + \bar{\nu}_j \leftrightarrow \nu_i + \bar{\nu}_j$ & $(p_1\!\cdot\! p_2)(p_3\!\cdot\! p_4)$ & $8\, T_{\nu_j}^4 T_{\nu_i}^4 \bigl[ T_{\nu_j} - T_{\nu_i} \bigr]$ \\
\end{tabular}
\end{ruledtabular}
\caption{Matrix element squared and energy transfer rates for neutrino decoupling processes in the Standard Model are summarized in Table \cite{EscuderoAbenza:2020cmq}. Here, $g_{L_e} = 1/2 + s_W^2$, $g_{R_e} = s_W^2$, $g_{L_{\mu,\tau}} = -1/2 + s_W^2$, $g_{R_{\mu,\tau}} = s_W^2$, $G_F = 1.166 \times 10^{-5}\,\text{GeV}^{-2}$ is the Fermi constant, and we use $s_W^2 = 0.231$.}
\label{tab:SMratesTab}
\end{table*}

\section{Energy Transfer Rates}
\label{app:rates}

For a generic $2\to2$ scattering process $1 + 2 \rightarrow 3 + 4$ with $T_1 = T_2 = T$ and $T_3 = T_4 = T'$, the energy transfer rate between the initial and final sectors in the Maxwell--Boltzmann approximation is \cite{EscuderoAbenza:2020cmq}:
\begin{align}
\frac{\delta \rho}{\delta t} ={}& \frac{g_1 g_2}{16\pi^4} \int_{s_{\min}}^{\infty} \dd s\, p_{12}^{2}\, \left(s + m_2^2 - m_1^2\right) \sigma(s) \nonumber \\
& \times \left[ T' K_{2}\!\left(\frac{\sqrt{s}}{T'}\right) - T K_{2}\!\left(\frac{\sqrt{s}}{T}\right) \right] , 
\label{eq:Tannrate} 
\end{align}
where $p_{12} = \sqrt{[s-(m_1+m_2)^2][s-(m_1-m_2)^2]} / (2\sqrt{s})$, $\sigma(s)$ is the spin-averaged cross section, and $s_{\min} = \text{Max}[(m_1+m_2)^2, (m_3+m_4)^2]$. Similarly for a decay $1 \rightarrow 2 + 3$ with daughter temperature $T_2 = T_3 = T$ and parent temperature $T_1$, the transfer rate is \cite{EscuderoAbenza:2020cmq}:
\begin{align}
\frac{\delta \rho}{\delta t} = \frac{g_1 m_1^3}{2\pi^2} \Gamma_{1\to 2 3} \left[ T K_{2}\!\left(\frac{m_1}{T}\right) - T_1 K_{2}\!\left(\frac{m_1}{T_1}\right) \right].
\label{eq:Tdecayrate}
\end{align}

\subsection{Standard Model Transfer Rates}
\label{app:SMrates}

In the SM, the relevant species are electrons, positrons, photons, and three active neutrino flavors at the MeV scale. The relevant interaction rates are summarized in the table~\ref{tab:SMratesTab}. 

Combining these SM $\nu-e$ rates in the following compact form \cite{EscuderoAbenza:2020cmq,Escudero:2019gzq}:
\begin{align}
\left(\frac{\delta \rho_{\nu_e - e}}{\delta t}\right)_{\text{SM}} &= \frac{4 G_F^2}{\pi^5} (g^2_{eL}+g^2_{eR})\, F_{\text{MB}}(T_{\gamma},T_{\nu_e}) \,,
\label{eq:delnui} \\[2pt]
\left(\frac{\delta \rho_{\nu_{\mu/\tau} - e}}{\delta t}\right)_{\text{SM}} &= \frac{4 G_F^2}{\pi^5}(g^2_{\mu L}+g^2_{\mu R})\, F_{\text{MB}}(T_\gamma,T_{\nu_\mu}) \,
\label{eq:SMrates}
\end{align}
where, $F_{\text{MB}}(T_1,T_2) = 32\, (T_1^9 - T_2^9) + 56\, T_1^4 T_2^4\, (T_1 - T_2) \ $. Thus the total SM $\nu - e$ transfer rate is
\begin{align}
\left(\frac{\delta \rho_{\nu - e}}{\delta t}\right)_{\text{SM}} &= \frac{2 G_F^2}{\pi^5} \left[ C^2_{V}+C^2_{A} + 2(\tilde{C}^2_{V}+\tilde{C}^2_{A}) \right] F_{\text{MB}}(T_{\gamma},T_{\nu})
\label{eq:delnuT} 
\end{align}
where $C_V = g_{L_e} + g_{R_e}$, $C_A = g_{L_e} - g_{R_e}$, $\tilde{C}_V = g_{L_{\mu,\tau}} + g_{R_{\mu,\tau}}$, and $\tilde{C}_A = g_{L_{\mu,\tau}} - g_{R_{\mu,\tau}}$.

\subsection{BSM Transfer Rates}
\label{app:BSMrates}

In our model, the additional energy-transfer rates involve $Z'$-mediated $\nu-e$ scattering, $Z' \leftrightarrow e^+e^-, \nu_i\bar\nu_i$, and $\chi\bar\chi \to e^+e^-$. The $Z'$-mediated $\nu-e$ rate with the interference term is obtained from \cite{Ghosh:2023ilw}. Since $T_{Z'} = T_\nu$, the net energy transfer from $Z' \leftrightarrow \nu_i\bar{\nu}_i$ vanishes, while the $Z' \leftrightarrow e^+e^-$ transfer rate is obtained from Eq.~\ref{eq:Tdecayrate} with $g_1 = g_{Z'} = 3$. Finally, for the dark matter annihilating into electron-positron pairs~\cite{EscuderoAbenza:2020cmq}, the energy transfer rate is
\begin{align}
\frac{\delta \rho_{\chi-e}}{\delta t} 
={}& \frac{g_{\chi}^2 m_{\chi}^5}{8\pi^4} \langle\sigma v\rangle_{\chi\bar\chi \to e^+e^-} \nonumber\\
&\times\left[ T_{\gamma}^2 K_2^2\!\left(\frac{m_{\chi}}{T_{\gamma}}\right)
- T_{\nu}^2 K_2^2\!\left(\frac{m_{\chi}}{T_{\nu}}\right) \right] .
\label{eq:wimprate2}
\end{align}
here $g_{\chi}=4$, is the DM internal degree of freedom.

\bibliographystyle{apsrev4-2}
\bibliography{main}

\end{document}